\documentclass[runningheads]{llncs}
\usepackage{graphicx}
\usepackage{hyperref}
\usepackage{amsmath,amssymb}

\begin{document}
%
\title{Deep PET/CT fusion with Dempster-Shafer theory for lymphoma segmentation\thanks{This work was supported by the China Scholarship Council (grant 201808331005). It was carried out in the framework of the Labex MS2T (Reference ANR-11-IDEX-0004-02)}}
\titlerunning{Deep PET/CT fusion with Dempster-Shafer theory}
%
\author{Ling Huang\inst{1} \and
Thierry Den{\oe}ux\inst{1,2} \and
David Tonnelet \inst{3} \and
Pierre Decazes \inst{3} \and
Su Ruan\inst{4}}
\authorrunning{L.Huang et al.}
\institute{Universit\'e de technologie de Compi\`egne, CNRS, Heudiasyc, Compi\`egne, France \\
\email{ling.huang@utc.fr} \and
Institut universitaire de France, Paris, France \and
CHB Hospital, Rouen, France \and
University of Rouen Normandy, Rouen, France}
\maketitle              
\begin{abstract}
Lymphoma detection and segmentation from whole-body Pos\-itron Emission Tomography/Computed Tomography (PET/CT) volumes are crucial for surgical indication and radiotherapy. Designing automatic segmentation methods capable of effectively exploiting the information from PET and CT as well as resolving their uncertainty remain a challenge. In this paper, we propose an lymphoma segmentation model using an UNet with an evidential PET/CT fusion layer. Single-modality volumes are trained separately to get initial segmentation maps and an evidential fusion layer is proposed to fuse the two pieces of evidence using Dempster-Shafer theory (DST). Moreover, a multi-task loss function is proposed: in addition to the use of the Dice loss for PET and CT segmentation, a loss function based on the concordance between the two segmentation is added to constrain the final segmentation. We evaluate our proposal on a database of polycentric PET/CT volumes of patients treated for lymphoma, delineated by the experts. Our method get accurate segmentation results with Dice score of 0.726, without any user interaction. Quantitative results show that our method is superior to the state-of-the-art methods.

\end{abstract}
\section{Introduction}

In the clinical diagnosis and radiotherapy planning of lymphoma, PET/CT scanning is an effective imaging tool for tumor segmentation. In PET volumes, the standardized uptake value (SUV) is widely used to locate and segment lymphomas because of its high sensitivity and specificity \cite{nestle2005comparison}. Moreover, CT is usually used in combination with PET because of its good representation of anatomical features. 

\begin{figure}
\includegraphics[width=\textwidth]{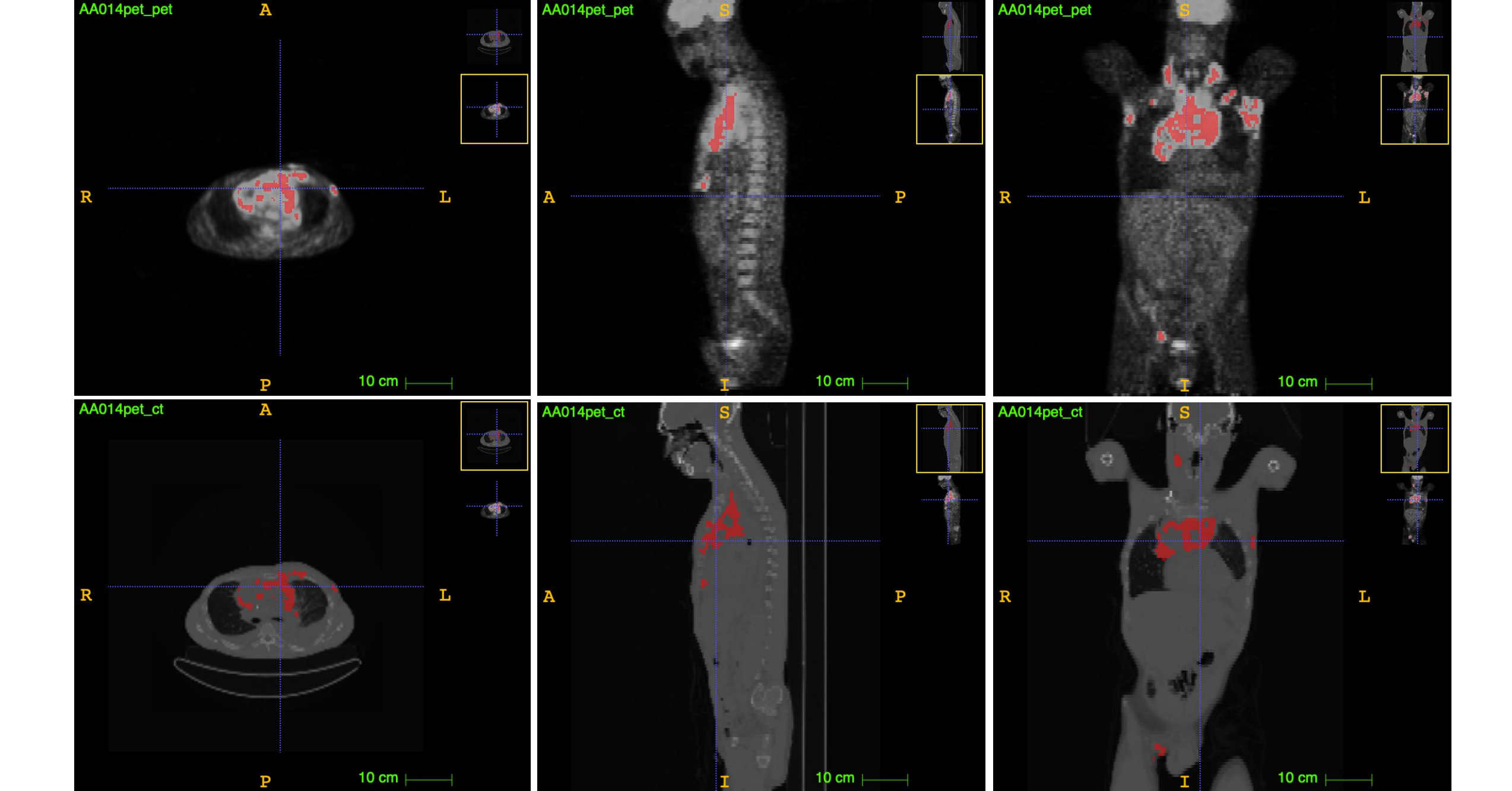}
\caption{Examples of PET and CT slices with lymphomas. The first and second row show PET and CT slices of one patient in axial, sagittal and coronal views, respectively.}
\label{fig1}
\end{figure}

Considering the multiplicity of lymphoma sites (sometimes more than 100) and the wide variation in distribution, shape, type and number of lymphomas, whole-body PET/CT lymphoma segmentation is still challenging although many segmentation methods have been proposed (see a lymphoma patient in Fig.~\ref{fig1}). Computer-aided methods for lymphoma segmentation can be classified into three main categories: SUV-threshold-based, region-growing-based and Convolutional-Neural-Network (CNN)-based \cite{li2019densex} methods.

For PET volumes, it is common to use fixed SUV thresholds to segment lymphoma. This kind of method is fast but it lacks flexibility in boundary delineation and requires clinicians to locate the region of interest. Region-growing-based methods have been proposed to overcome the limitation of boundary delineation of SUV-based methods by taking texture and shape information into account. The common idea is to set the volumes with initializing seeds automatically for segmenting lymphomas \cite{li2008novel}\cite{desbordes20143d}. However, these methods still require to region of interest manually and are time-consuming.

In recent years, CNNs have achieved great success in computer vision tasks \cite{hershey2017cnn}\cite{bochkovskiy2020yolov4}, as well as in the medical domain. The Fully Connected Network (FCN) model \cite{long2015fully} is the first fully convolutional neural network that could be trained end-to-end for pixel-wise classification. UNet \cite{ronnebergerconvolutional}, a successful modification and extension of FCN, has become the most popular network for medical image segmentation. Based on UNet, many extensions and optimizations have been proposed, such as Deep3D-UNet \cite{zhu2018anatomynet}, attention-UNet \cite{oktay2018attention}, etc. In \cite{hu2020lymphoma}, Hu et al. propose a Conv3D-based multi-view PET image fusion strategy for lymphoma segmentation. In this approach, 2D and 3D segmentation are performed separately and the results are fused by a Conv3D layer. In \cite{li2019densex}, Li et al. propose a whole-body PET/CT lymphoma segmentation method that fuses CT  and PET slices by concatenating them before training; the method is based on a two-flow architecture (segmentation flow and reconstruction flow). Using a similar approach, Blanc-Durand et al. propose a CNN-based segmentation network for diffuse large B cell lymphoma segmentation by concatenating PET and CT as two channel inputs \cite{blanc2020fully}.

It should be noted that the effective fusion of multi-modality information is of great importance in the medical domain. A single-modality image often does not contain enough information and is often tainted with uncertainty. This is why physicians always use PET/CT volumes together for lymphoma segmentation and radiotherapy. 
Using CNNs, researchers have mainly adopted probabilistic approaches to data fusion, which can be classified into three strategies \cite{zhou2019review}: image-level, feature-level  and decision-level fusion. 
However, probabilistic fusion is unable to effectively manage conflicts that occur when the same voxel is labeled with two different labels by CT and TEP. Dempster-Shafer theory (DST) \cite{dempster1967upper}\cite{shafer1976mathematical}, also known as belief function theory or evidence theory, is a formal framework for information modeling, evidence combination and decision-making with uncertain or imprecise information. Despite the low resolution and contrast of medical images, DST's high ability to describe uncertainty allows us to represent evidence more faithfully than using probabilistic approaches. Researchers from the medical image community have started to actively investigate the use of DST for handling uncertain, imprecision sources of information in different medical tasks, such as medical image retrieval \cite{sundararajan2019deep}, lesion segmentation \cite{lian2018joint}, etc.

In this work, we propose a DST-based PET/CT image fusion model for 3D lymphoma segmentation. To our knowledge, this is the first multi-modality PET/CT volume fusion method using a CNN and DST. The main contributions of this work are (1) a CNN architecture with a DST-based fusion layer that effectively handles uncertainty and conflict when combining PET and CT information; (2) a multi-task loss function making it possible to optimize different segmentation tasks and to increase segmentation accuracy; (3) a 3D segmentation model with end-to-end training for whole-body lymphoma segmentation.

\section{Methods}

\subsection{Network Architecture}
\begin{figure}
\includegraphics[width=\textwidth]{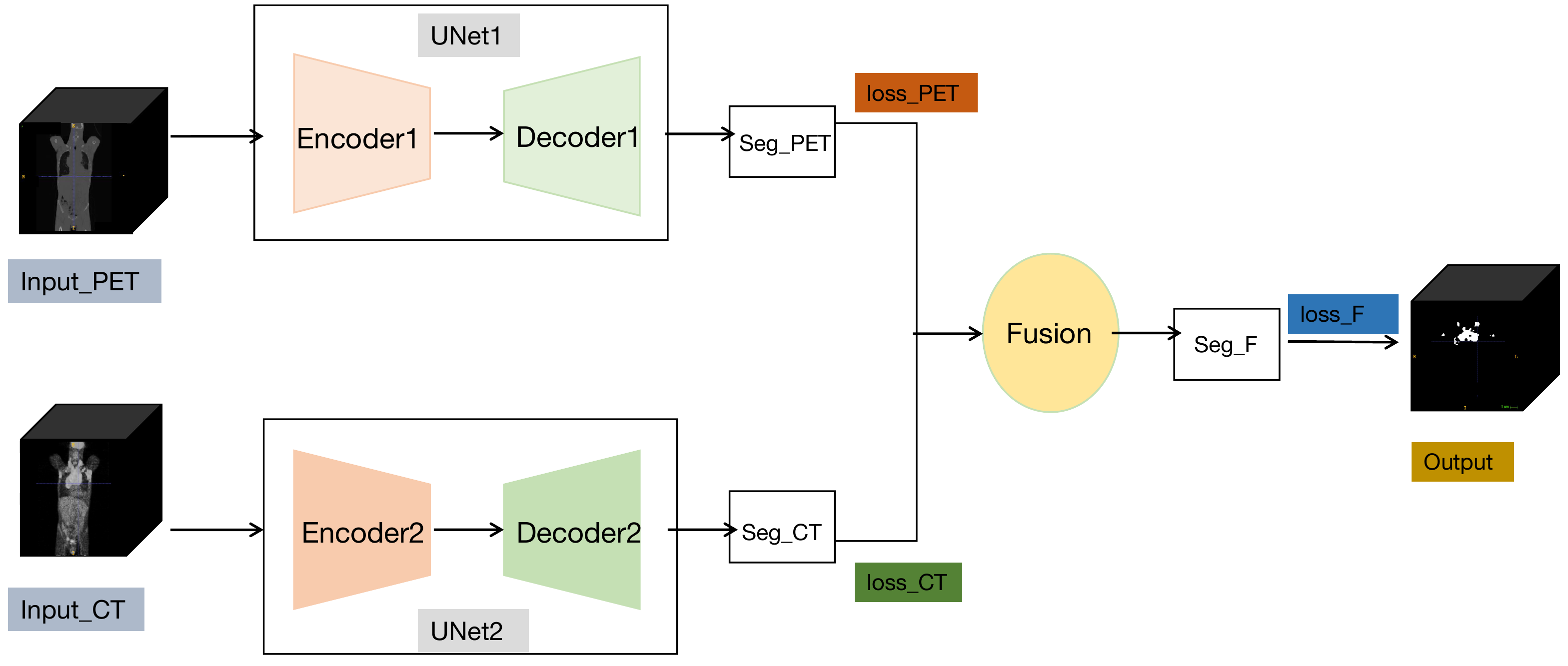}
\caption{The global segmentation framework.}
\label{fig2}
\end{figure}

Fig.~\ref{fig2} shows the workflow of the multi-modality fusion-based lymphoma segmentation framework. It is composed of two modified encoder-decoder (Unet) modules and an evidential fusion layer. To reduce computation cost, we reduce the number of convolution filters of Unet from $(16, 32, 64, 128, 256)$ to $(8, 16, 32, 64, 128)$ to get a ``slim UNet''. A fusion layer is constructed based on DST, which will be explained in Section 2.3. Two modality images: PET and CT are taken as inputs to our framework. We first feed the prepossessed PET volume into UNet1 and the prepossessed CT volume into UNet2 to independently compute their segmentation probability maps $\textsf{seg}_{PET}$ and $\textsf{seg}_{CT}$. These two 3D maps are then transferred to the fusion layer, which computes an evidential segmentation map $\textsf{seg}_F$. For training, a multi-task loss function is proposed to minimize the Dice loss and mean square loss between three segmentation maps and masks, as explained in Section 2.4.

\subsection{Evidential Fusion Strategy}
 Let $\Omega =\{\omega _{1}, \omega _{2}, ..., \omega _{K}\} $ be a finite set of possible answers to some question, one and only one of which is true. Evidence about that question can be represented by a mapping $m$ from $2^\Omega$ to $[0,1]$, called a mass function, such that $\sum _{A\subseteq \Omega} m(A)=1$. For any hypothesis $A\subseteq\Omega$, the quantity $m(A)$ represents a share of a unit mass of belief allocated to the hypothesis that the truth is in $A$, and which cannot be allocated to any strict subset of $A$ based on the available evidence. Two mass functions $m_{1}$ and $m_{2}$ representing two independent items of evidence can be combined by Dempster's rule \cite{shafer1976mathematical} defined as 
\begin{equation}
\label{eq:dempster}
    (m_{1}\oplus m_{2})(A)=\frac{1}{1-\kappa }\sum _{B\cap C=A}m_{1}(B)m_{2}(C),
\end{equation}
for all $A\subseteq \Omega, A\neq \emptyset$, and $(m_{1}\oplus m_{2})(\emptyset)=0$. In Eq. \eqref{eq:dempster}, $\kappa$ represents the degree conflict between $m_{1}$ and $m_{2}$ defined as
\begin{equation}
\label{eq:conflict}
    \kappa=\sum _{B\cap C=\emptyset}m_{1}(B)m_{2}(C).
\end{equation}

If all focal sets of $m$ are singletons and $m$ is said to be Bayesian, it is equivalent to a probability distribution. In our case, the lymphoma segmentation task can be considered as a two-class classification task; the set of possibilities is $\Omega=\{0,1\}$ where 1 and 0 stand, respectively, for presence and absence of lymphoma in a given voxel. Two Bayesian mass functions $m_{\textsf{PET}}$ and $m_{\textsf{CT}}$
can be obtained from two probability segmentation maps maps $\textsf{seg}_{PET}$ and $\textsf{seg}_{CT}$. They are aggregated by Dempster's rule \eqref{eq:dempster}.

The two segmentation results obtained by PET and CT cannot easily classify boundary voxels due to low signal-to-noise ratio and low contrast, resulting in segmentation uncertainty. Moreover, when $\textsf{seg}_{PET}$ and $\textsf{seg}_{CT}$ assign the same voxel to different classes, there is a high conflict between the two segmentation results. It is difficult to solve these problems by classical fusion methods. In this work, the proposed evidential fusion layer allows us to solve these problems thanks to Dempster's rule. To be specific, for a given voxel, if $\textsf{seg}_{PET}$ gives a probability of 0.26 that it belongs to lymphoma, and $\textsf{seg}_{CT}$ gives a probability of 0.85 that it belongs to lymphoma, the two segmentation results are contradictory and it is unreasonable to simply fuse them by linear combination or majority voting. In our evidential fusion layer, we can reassign the probability that the voxel belongs to lymphoma as $0.67$ with Eqs. \eqref{eq:dempster}-\eqref{eq:conflict}. This rule takes the conflict into consideration and yields more reliable segmentation results. 

\subsection{Multi-task loss function}



Lymphomas segmentation task show sub-optimal performance in CT, which may leads to misleading fusion results in our model. In order to avoid the problem, a multi-task loss function here is proposed to exploit all the available information during training and increase segmentation accuracy. Since lymphomas are visible on both PET and CT, even though they do not have exactly the same shape in these two volumes, they should overlap as much as possible. Here we set PET mask as ground truth for both PET and CT to constrains the overlap rate between PET and CT.

As shown in Fig. \ref{fig1}, we define three  loss functions: $\textsf{loss}_{CT}$,  $\textsf{loss}_{PET}$ and  $\textsf{loss}_F$ measuring the discrepancy between ground truth and, respectively, CT segmentation maps, PET segmentation maps, and the final segmentation output.
For $\textsf{loss}_{CT}$ and $\textsf{loss}_{PET}$, we use Dice loss,
\begin{equation}
        \textsf{loss}_{PET}=1-\frac{2* \sum_{v=1}^{V} S_{1}^v G^v}{ \sum_{v=1}^{V} S_{1}^v+ \sum_{v=1}^{V} G^v} ,
\end{equation}
\begin{equation}
    \textsf{loss}_{CT}=1-\frac{2* \sum_{v=1}^{V} S_{2}^v G^v}{ \sum_{v=1}^{V} S_{2}^v+ \sum_{v=1}^{V} G^v},
\end{equation}
where $V$ is the number of voxels of segmentation outputs, $S_1$ and $S_2$ are the PET and CT segmentation outputs and $G$ is the ground truth of lymphoma in PET. For $\textsf{loss}_F$, we use mean square loss,
\begin{equation}
      \textsf{loss}_{F}= \sum_{v=1}^{V}(S_{f}^v-G^v)^{2},
\end{equation}
where $S_f$ is the final segmentation output. The multi-task loss function is defined as
\begin{equation}
\label{loss}
\textsf{loss}_{all}=0.75*\textsf{loss}_{CT}+0.25*\textsf{loss}_{PET}+\textsf{loss}_{F}.
\end{equation}
Here we set the weight of ${loss}_{CT}$ as 0.75 and the weight of ${loss}_{PET}$ as 0.25 to enable the model learn more from hard example (CT). 

\subsection{Implementation Details}
All methods were implemented in Python using Tensorflow framework and were trained and tested on a desktop with a 2.20GHz Intel(R) Xeon(R) CPU E5-2698 v4 and a Tesla V100-SXM2 graphics card with 32 GB GPU memory. 

\section{Experiments and Analysis}

\subsection{Dataset and Preprocessing}

The experimental dataset consists of 173 labeled cases of real PET/CT volumes, whose labels indicate the ground truth of lymphoma. All PET/CT data were acquired from the Henri Becquerel hospital. The study was approved as a retrospective study by the Henri Becquerel Center Institutional Review Board. All patients' information were de-identified and anonymized prior to analysis. The size of the CT volumes and the corresponding masks vary from $267\times 512\times512$ to $478\times 512\times512$ and their spatial resolution varied from $0.97 \times 0.97\times 2 $ $mm^3$ to $1.36 \times 1.37 \times 5$ $mm^3$. The size of the PET volumes and the corresponding masks vary from $276\times 144\times144$ to $407\times 256\times256$ and their spatial resolution varied from $5.3 \times 5.3 \times 2$ $mm^3$ to $2.73 \times 2.73 \times 3.27 $ $mm^3$. For preprossessing, we first registered the CT volumes and matched them with PET volumes. Then we resized PET, CT and mask volumes into size $128\times 256\times256$. Both PET and CT volumes were normalized into the standard distribution by the linear standardization method. Two kinds of data augmentation were applied here to enrich the training data: deformation and affine transformation. We used 80\% of the data for training, 10\% for validation and 10\% for testing. 
The Dice score, Precision and Recall were used  to evaluate the segmentation results of lymphoma at the voxel level.

\subsection{Results and Discussion}
\begin{figure}
\includegraphics[width=\textwidth]{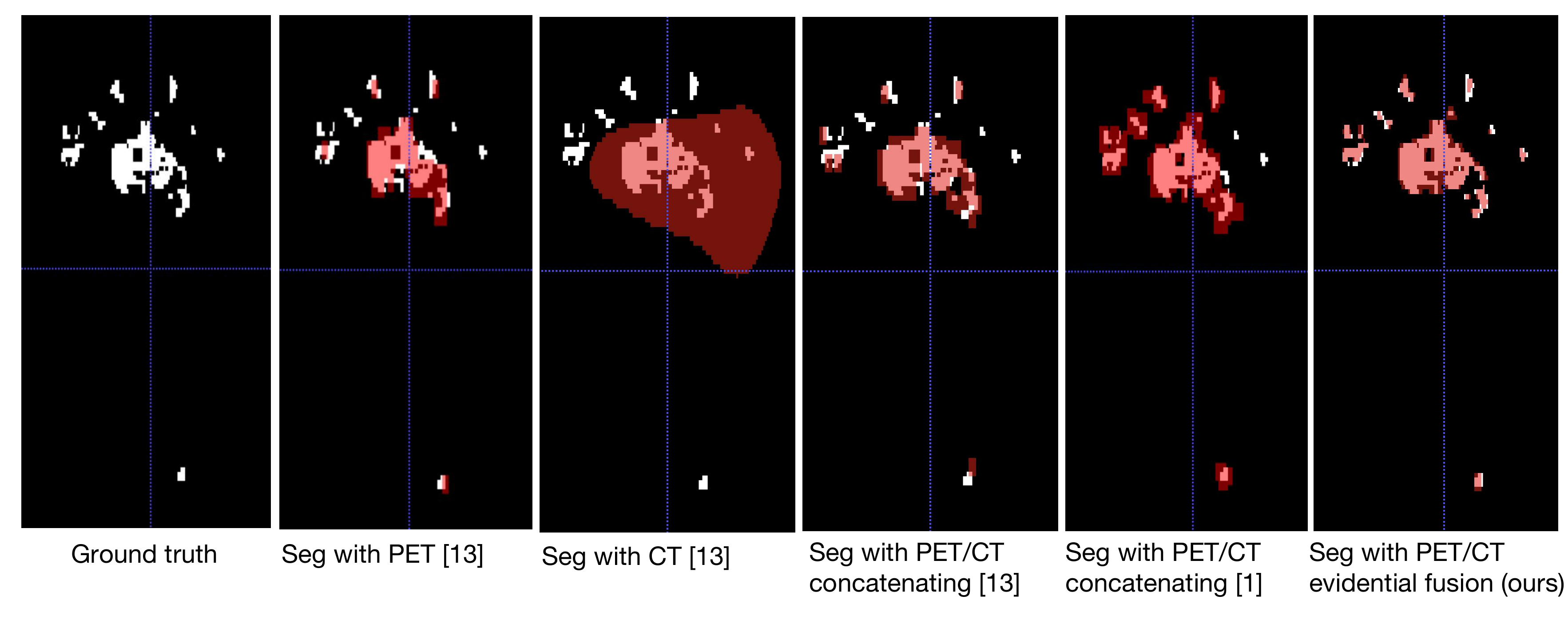}
\caption{Comparison of segmentation results of CT, PET, PET/CT concatenation \cite{ronnebergerconvolutional}, PET/CT concatenation with \cite{blanc2020fully} and PET/CT evidential fusion. The ground truth is marked in white and segmentation results are marked in red.}
\label{fig3}
\end{figure}
The quantitative results of the experiments are represented in Table~\ref{tab1}. UNet is the baseline method. We first tested the segmentation performance when using only PET or CT volume. Then we concatenate PET and CT volumes as inputs of the UNet. This is an input-level fusion method. Finally, we used two independent UNets for PET and CT, respectively. The fusion was carried out at decision level by the proposed evidential fusion layer. Compared with mono-modality input, our proposal shows an obvious advantage. Compared with the concatenate-based fusion method, our proposal has 3\%, 4\%, 6\% increase in Dice score, Precision and Recall, respectively. Fig.~\ref{fig3} shows the comparison results between the above four input modalities. We can see from the figure that mono-modality input does not allow us to segment small regions, especially for CT. The concatenation of PET and CT solves this problem to some extent \cite{blanc2020fully}\cite{ronnebergerconvolutional}, but the result is still not ideal. The evidential fusion of PET and CT achieves better segmentation result, especially for the small lymphomas located at the top of the images.

\begin{table}
\caption{Performance comparison with the baseline methods on the test set.}
\label{tab1}
\begin{tabular}{|l|l|l|l|l|}
\hline
Models  &Input Modality & Dice score &Precision &Recall\\
\hline
UNet (mono-input)  &PET & 0.67$\pm$0.02 & 0.74$\pm$0.05 & 0.65$\pm$ 0.03 \\
UNet (mono-input) & CT & 0.49$\pm$ 0.05 & 0.71$\pm$ 0.02 & 0.45$\pm$ 0.02\\
UNet (concatenate-based fusion) & PET+CT &0.69$\pm$ 0.01 & 0.67$\pm$ 0.01& 0.74$\pm$ 0.05\\
EUnet (evidential fusion)& PET+CT & \textbf{0.72}$\pm$ 0.04&
\textbf{0.71}$\pm$ 0.06 & \textbf{0.80}$\pm$ 0.05 \\
\hline
Zeng et al. \cite{zeng20173d} &PET+CT &0.68±0.02&0.68±0.01&0.68±0.01\\
Hu et al. \cite{hu2020lymphoma}&PET+CT&
0.66±0.03&0.71±0.044&0.67±0.03\\
Blanc et al.\cite{blanc2020fully}&PET+CT&\textbf{0.73}±0.20& \textbf{ 0.75}±0.22& \textbf{0.83}±0.17\\
\hline

\end{tabular}
\end{table}

Since there is no public lymphoma dataset available now, the quantitative comparison with other state-of-the-art methods is difficult, because different datasets and different lymphoma types are segmented. However, we still compare our method with the state-of-the-art in Table~\ref{tab1}. In \cite{hu2020lymphoma}, Hu et al. achieve 0.66±0.03 Dice score on a dataset of 109 lymphoma patients with a convolution fusion with 2D slices and 3D volumes. In \cite{blanc2020fully}, Blanc-Durand et al. report 0.73±0.20 for Dice score based on the training on 511 lymphoma patients. Though the authors report comparable results with ours, the performance of our model is more stable and needs less training data. We also test the method of \cite{blanc2020fully} with our dataset and achieve 0.64±0.02 in Dice score, where the priory of our proposal is obvious. A qualitative comparison with our model is shown in Fig.~\ref{fig3}. See the fifth image from Fig.~\ref{fig3}, their model show sign of slight overfitting but is visually acceptable and our model show visually correct segmentation results. Generally speaking, our model yields better results than state-of-the-art methods with DST-based evidential fusion.

\begin{figure}
\includegraphics[width=\textwidth]{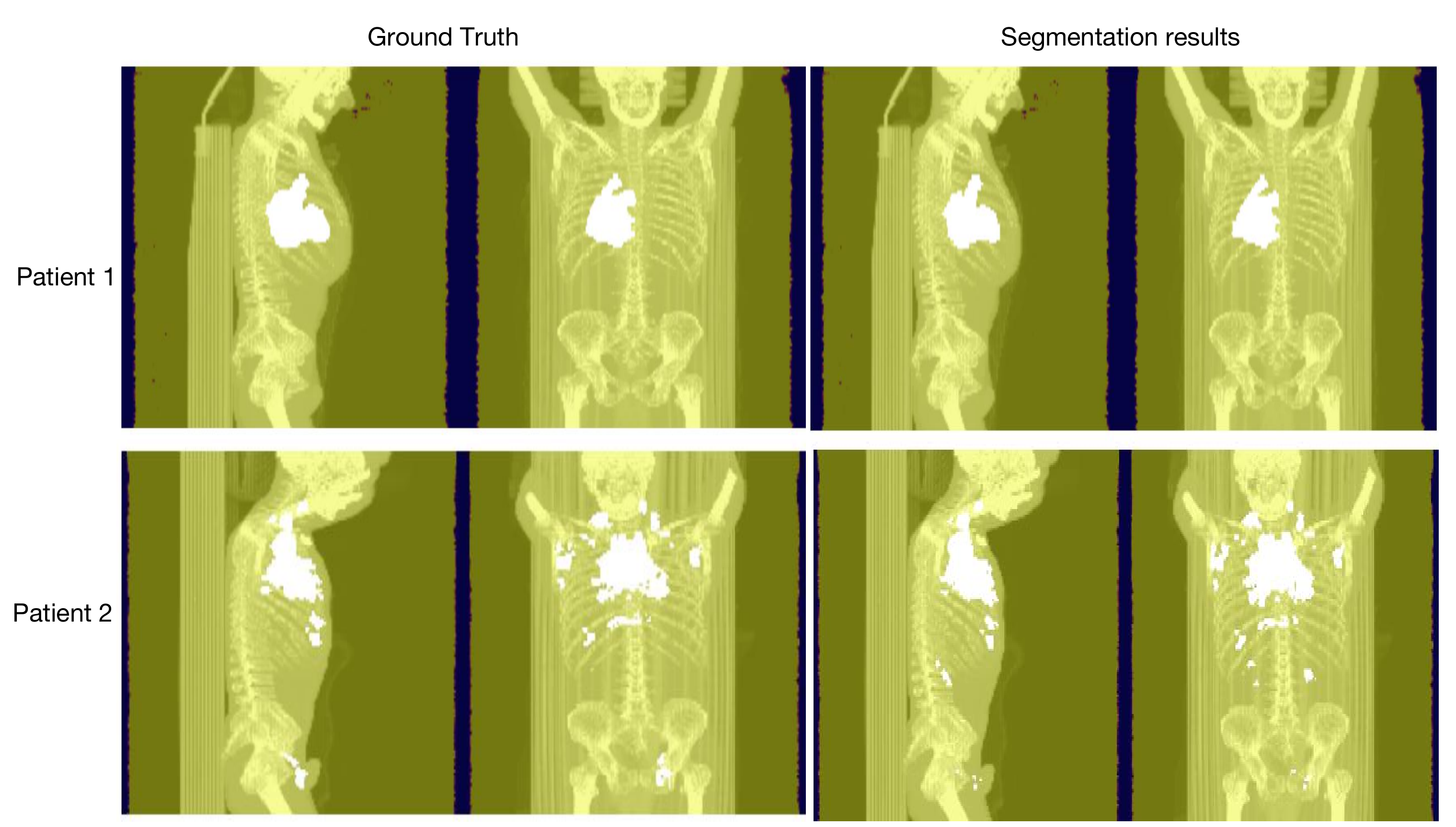}
\caption{Comparison results between two different patients by using our EUNet model. From left to right, each column represents the corresponding segmentation results and the ground truth. The lymphomas are showed in white.}
\label{fig4}
\end{figure}

A qualitative comparison is shown in 
Fig.~\ref{fig4}. Here we overlap PET and CT in the same image and paste the segmented mask in this image. Column 1 shows one slice of patient 1 in which a big lymphoma is present. Our model can segment it correctly. Column 2 shows a slice of patient 2 in which multiple lymphomas are present. They are more difficult to segment because some of them are very small. However, our model can still obtain satisfactory results.

\section{Conclusion}
In this work, a DST-based deep multi-modality medical images fusion strategy has been proposed to deal with the problem of uncertainty and conflict within a deep convolutional neural network. A two-branch segmentation module first processes PET/CT volumes separately. Then the two segmentation maps are fused by the evidential fusion layer. Qualitative and quantitative evaluation results are promising as compared to the baseline and state-of-the-art methods. Future research will aim at improving the network architecture to better exploit the potential of DST to represent and combine uncertain information. 



%
%

%
%
%
 \bibliographystyle{splncs04}
 \bibliography{ref}
%




\end{document}